\documentclass[a4paper,fleqn,usenatbib]{mnras}
\usepackage{newtxtext,newtxmath}
\usepackage[T1]{fontenc}
\usepackage{ae,aecompl}
\usepackage{graphicx}	
\usepackage{amssymb}	
\usepackage{graphics}
\usepackage{hyperref}

\def\asec{\ifmmode ^{\prime\prime}\else$^{\prime\prime}$\fi}

\def\it{\sl}
\def\degs{\ifmmode ^{\circ}\else$^{\circ}$\fi}
\def\amin{\ifmmode ^{\prime}\else$^{\prime}$\fi}
\def\asec{\ifmmode ^{\prime\prime}\else$^{\prime\prime}$\fi}
\def\farcs{\hbox{$.\!\!^{\prime\prime}$}}  
\def\h{$^{\rm h}$}
\def\m{$^{\rm m}$}

\def\psr{PSR J1826$-$1256}
\def\pmo{proper motion}
\def\j1826{J1826$-$1256}
\def\xmm{XMM-\textit{Newton}}
\def\chan{\textit{Chandra}}
\def\fermi{\textit{Fermi}}
\def\degs{\ifmmode ^{\circ}\else$^{\circ}$\fi}
\def\amin{\ifmmode ^{\prime}\else$^{\prime}$\fi}

\def\eqalign#1{\null\,\vcenter{\openup1\jot \m@th
   \ialign{\strut\hfil$\displaystyle{##}$&$\displaystyle{{}##}$\hfil
   \crcr#1\crcr}}\,}

\usepackage[normalem]{ulem}  

\title[X-ray studies of PSR \j1826]{X-ray studies of the gamma-ray pulsar \j1826\ and its pulsar wind nebula with \chan\ and \xmm}

\author[A. V. Karpova, D. A. Zyuzin and Y. A. Shibanov]{
Anna V. Karpova$^{1}$\thanks{E-mail: annakarpova1989@gmail.com},
Dmitry A. Zyuzin$^{1}$\thanks{E-mail: da.zyuzin@gmail.com}
and Yuriy A. Shibanov$^{1}$
\\
$^{1}$Ioffe Institute, Politekhnicheskaya 26, St. Petersburg, 194021, Russia
}

\date{Accepted XXX. Received YYY; in original form ZZZ}

\pubyear{2019}
\begin{document}
\label{firstpage}
\pagerange{\pageref{firstpage}--\pageref{lastpage}}
\maketitle

\begin{abstract}
We have analyzed archival \xmm\ and \chan\ observations of 
the $\gamma$-ray radio-quiet pulsar \j1826\ and its pulsar wind nebula.
The pulsar spectrum can be described by a power-law model with
a photon index $\Gamma\approx1$.
We find that the nebular spectrum softens with increasing distance
from the pulsar, implying synchrotron cooling. 
The empirical interstellar absorption--distance relation gives
a distance of $\approx3.5$ kpc to \j1826.
We also discuss the nebula geometry and 
association between the pulsar, the very high energy source HESS J1826$-$130,
the supernova remnant candidate G18.45$-$0.42 and the open star cluster Bica~3. 
\end{abstract}



\section{Introduction}
\label{intro}
The \fermi\ observatory opened a new window 
in the studies of $\gamma$-ray emission from compact objects.
To date, the Large Area Telescope (LAT) onboard \fermi\ 
has detected more than 200 rotation powered pulsars\footnote{\url{https://confluence.slac.stanford.edu/display/GLAMCOG/Public+List+of+LAT-Detected+Gamma-Ray+Pulsars}}. 
Multiwavelength observations of pulsars allow researchers 
to study the pulsar emission geometry,
particle acceleration mechanisms in the pulsar magnetosphere and
pulsar emission efficiencies as a function of energy 
\citep[e.g.][]{takata2017}.
They are also important for investigation of the thermal emission 
from pulsar surfaces \citep[e.g.][]{kargaltsev2007}.
Since more than a quarter of \fermi\ pulsars is radio-quiet,
X-ray observations play a key role in understanding their 
properties \citep[e.g.][]{marelli2013, marelli2015}.
Investigations in this band
can reveal their pulsar wind nebulae (PWNe) and associated supernova remnants (SNRs).
Studies of the latter objects, in turn, provide additional information 
about pulsar parameters (e.g. ages, distances, proper motions, geometries) 
and interaction of relativistic pulsar winds with the ambient medium
\citep[e.g.,][]{gaensler2006,kargaltsev2017}. 

The young and energetic radio-quiet PSR \j1826\  
was one of the first pulsars discovered in $\gamma$-rays
using blind frequency searches in \fermi\ data \citep{abdo2009}.
This is one of the brightest radio-quiet $\gamma$-ray pulsars 
listed in the last \fermi\ catalog 
\citep[3FGL J1826.1$-$1256;][]{acero2015}.
It has a period $P=110.2$ ms, a spin-down luminosity $\dot{E}=3.6\times10^{36}$ erg~s$^{-1}$, 
a characteristic age $\tau_c=14$ kyr and a surface magnetic field $B=3.7\times10^{12}$ G. 
Judging by these parameters the pulsar
belongs to a group of `Vela-like' pulsars. 
The distance (`pseudo'-distance) to \psr\ was estimated to be 
$\sim1.2$~kpc \citep{marelliphd} using 
the empirical relation between the distance 
and the $\gamma$-ray flux above 100 MeV \citep{SazParkinson2010}.
This estimate is known to be very uncertain.
\citet{ray2011} performed precise $\gamma$-ray timing analysis
improving the accuracy of the pulsar's \fermi\ position.
Its coincidence with an X-ray point-like source
previously detected with \chan\ \citep{roberts2007}
implied that the latter is the \psr\ counterpart.
Detection in the \xmm\ data of X-ray pulsations with the pulsar 
period from the point source \citep{Li2018apjl} confirmed its pulsar nature.
Analyzing \chan\ data, \citet{marelliphd} showed that \psr\ has a flat X-ray spectrum
described by a power law (PL) with the photon index $\Gamma=0.79\pm0.39$.

Before the pulsar's discovery, the \textit{Advanced Satellite for Cosmology and Astrophysics 
(ASCA)} observatory 
revealed a $\sim 15$ arcmin X-ray nebula AX J1826.1$-$1300, 
presumably a PWN, detected within the error-box of a bright 
previously unidentified $\gamma$-ray source 3EG J1826$-$1302 \citep{roberts2001}. 
Observations with \chan\ \citep{roberts2007} 
resolved in the nebula beside the pulsar, a faint, remarkably long trail-like PWN (G18.5$-$0.4)  
and a stellar cluster Bica~3\footnote{Bica~3 can be found in the New Optically Visible Open Clusters and Candidates Catalog \citep[OPENCLUST;][]{dias2002}; \url{https://heasarc.gsfc.nasa.gov/W3Browse/all/openclust.html}}. 
G18.5$-$0.4 is connected to the pulsar and extended by $\sim 4$ arcmin south-west from it.
Owing to its shape, it was referred to as the Eel nebula \citep{roberts2007}. 
\citet{marelliphd} found that its X-ray emission 
extracted within 20 arcsec from the pulsar has a PL spectrum 
with $\Gamma=0.86\pm0.39$, which is marginally steeper than that of  \psr. 
\citet{roberts2007} also noted, that there is a 90 cm radio emission,
which may be related to the Eel by its morphology, and the absence 
of associated mid-infrared emission implies a non-thermal origin.
Using \fermi\ data, \citet{ackermann2011} performed the maximum
likelihood spectral fits for the \psr\ off-pulse emission and did not 
find any signature of the PWN  in the GeV range.

At the same time, Eel together with \psr\ overlap with the extended TeV source 
HESS J1826$-$130 which was previously considered to be a part of the brighter 
TeV PWN HESS J1825$-$137 \citep{anguner2017o,hesscollaboration2018}. 
The source has a very hard spectrum similar to that of  
the Vela X PWN, which can be produced by uncooled electrons  with the spectral 
index close to 2.0 and a cut-off energy at around 70 TeV generated 
by the PSR$+$PWN system \citep{anguner2017o}. 
The TeV source was proposed to be associated with the Eel 
\citep{roberts2007}, although a firm conclusion was not possible, in part due to the  
uncertainty of the distance to the pulsar \citep{hesscollaboration2018}.  

HESS J1826$-$130 also partially overlaps with the SNR G18.6$-$0.2 \citep{brogan2006} 
which may indicate their association \citep{hesscollaboration2018}.
However, the latter is unlikely since this SNR has a significantly smaller size 
and a large offset from the TeV source center.

In the radio and mid-infrared, \citet{anderson2017} detected 
a shell-like SNR candidate G18.45$-$0.42. 
We found that its position and size are well compatible with the Eel and  
HESS J1826$-$130, making its association with these objects possible.

Here we report the simultaneous X-ray analysis of archival \xmm\ and \chan\ 
observations of \psr\ and its PWN. 
The details of observations and imaging are described in Section~\ref{sec:data}. 
The spectral analysis is presented in Section~\ref{sec:spectra}.
We discuss the results in Section~\ref{sec:discussion}. 
Based on the interstellar absorption versus distance relation we obtain a new distance estimation to 
the PSR$+$PWN system of $\approx 3.5$~kpc. 
We consider G18.45$-$0.42 as the best SNR candidate  
for possible association  with the \psr\ and its PWN. 
The summary is given in Section~\ref{sec:summary}. 

\begin{figure*}
\begin{minipage}[h]{0.495\linewidth}
\center{\includegraphics[width=0.97\linewidth,clip]{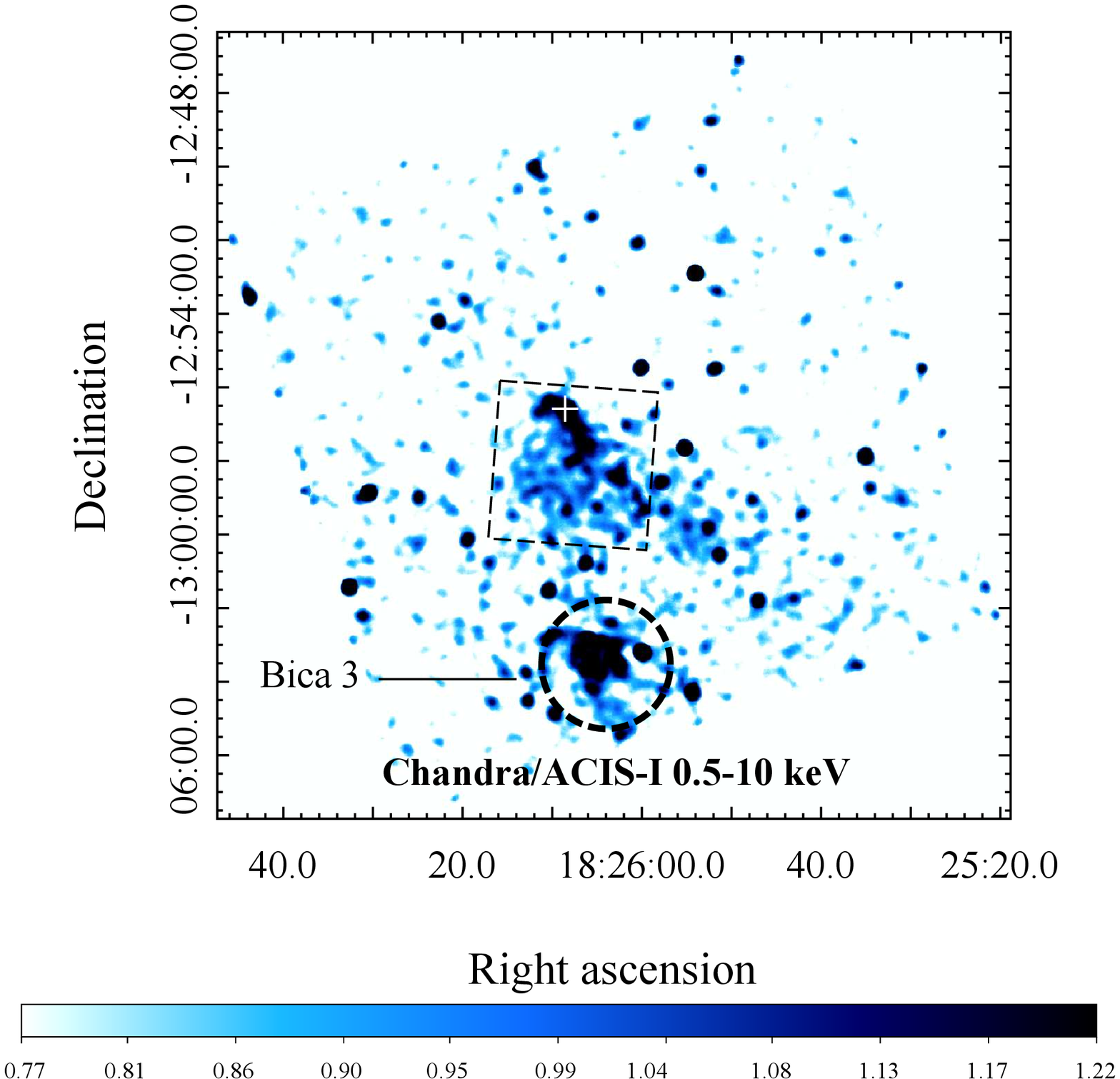}}
\end{minipage}
\begin{minipage}[h]{0.495\linewidth}
\center{\includegraphics[width=0.87\linewidth,clip]{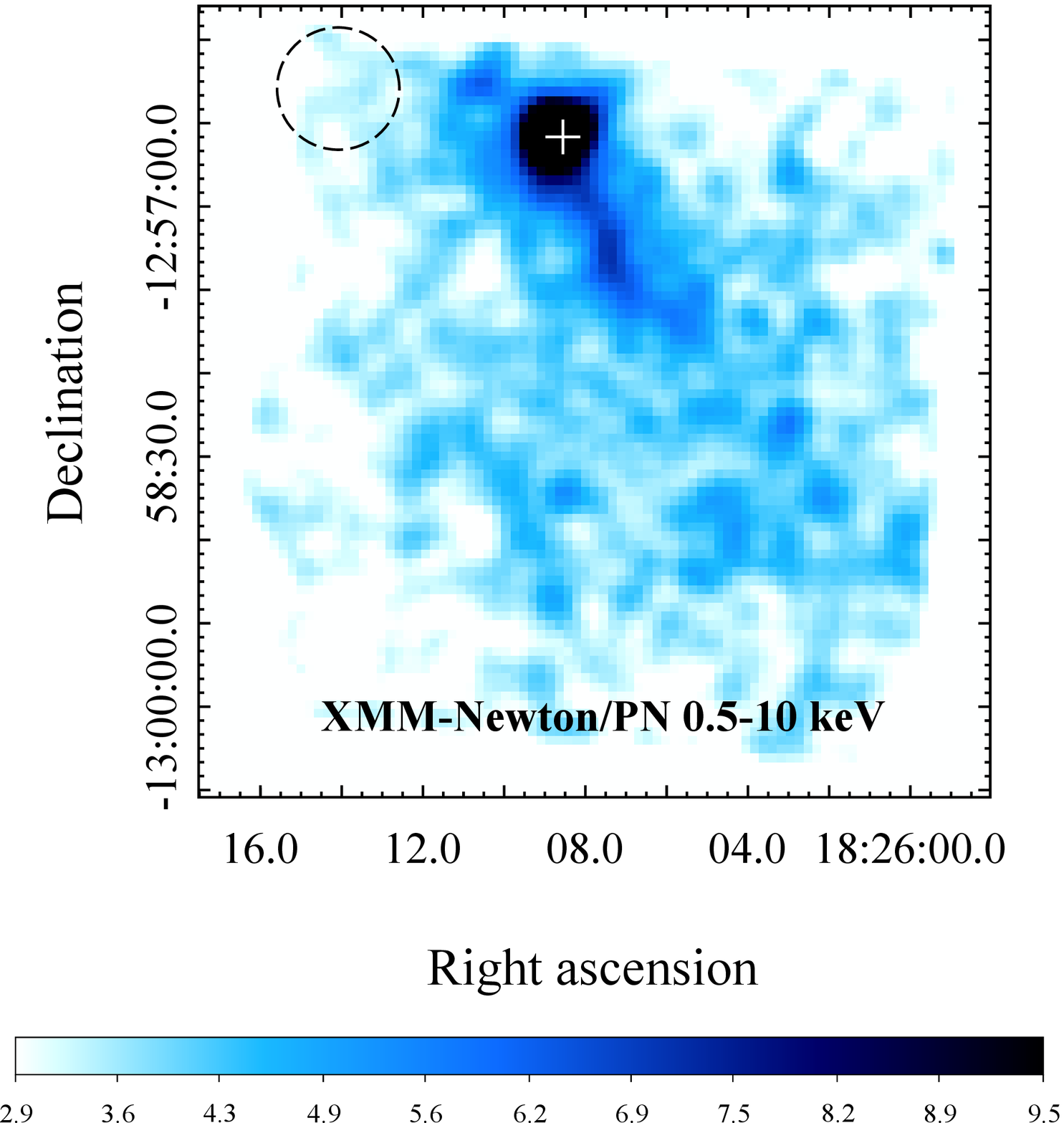}}
\end{minipage}
\begin{minipage}[h]{0.497\linewidth}
\center{\includegraphics[width=0.94\linewidth,clip]{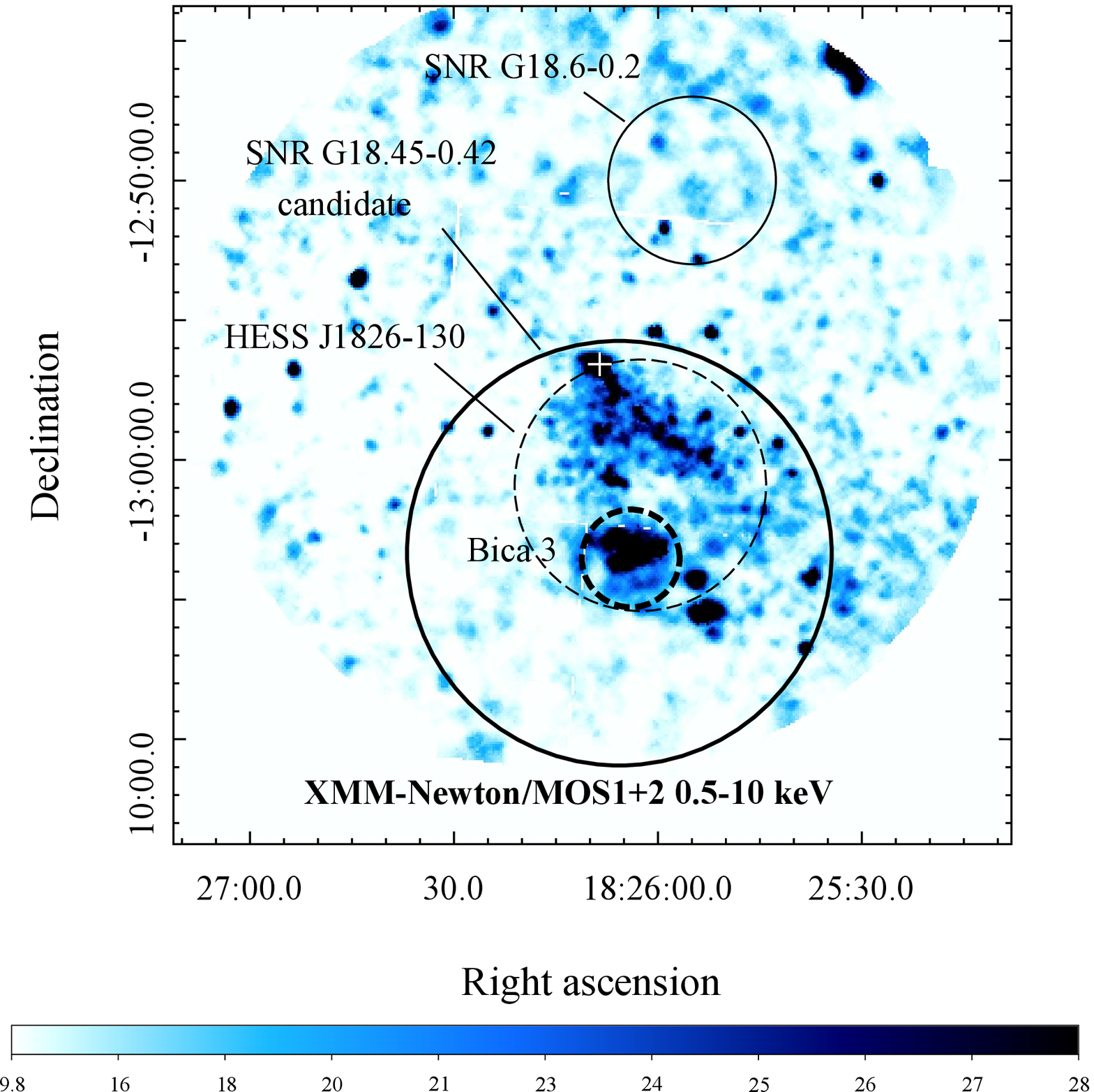}}
\end{minipage}
\begin{minipage}[h]{0.493\linewidth}
\center{\includegraphics[width=0.93\linewidth,clip]{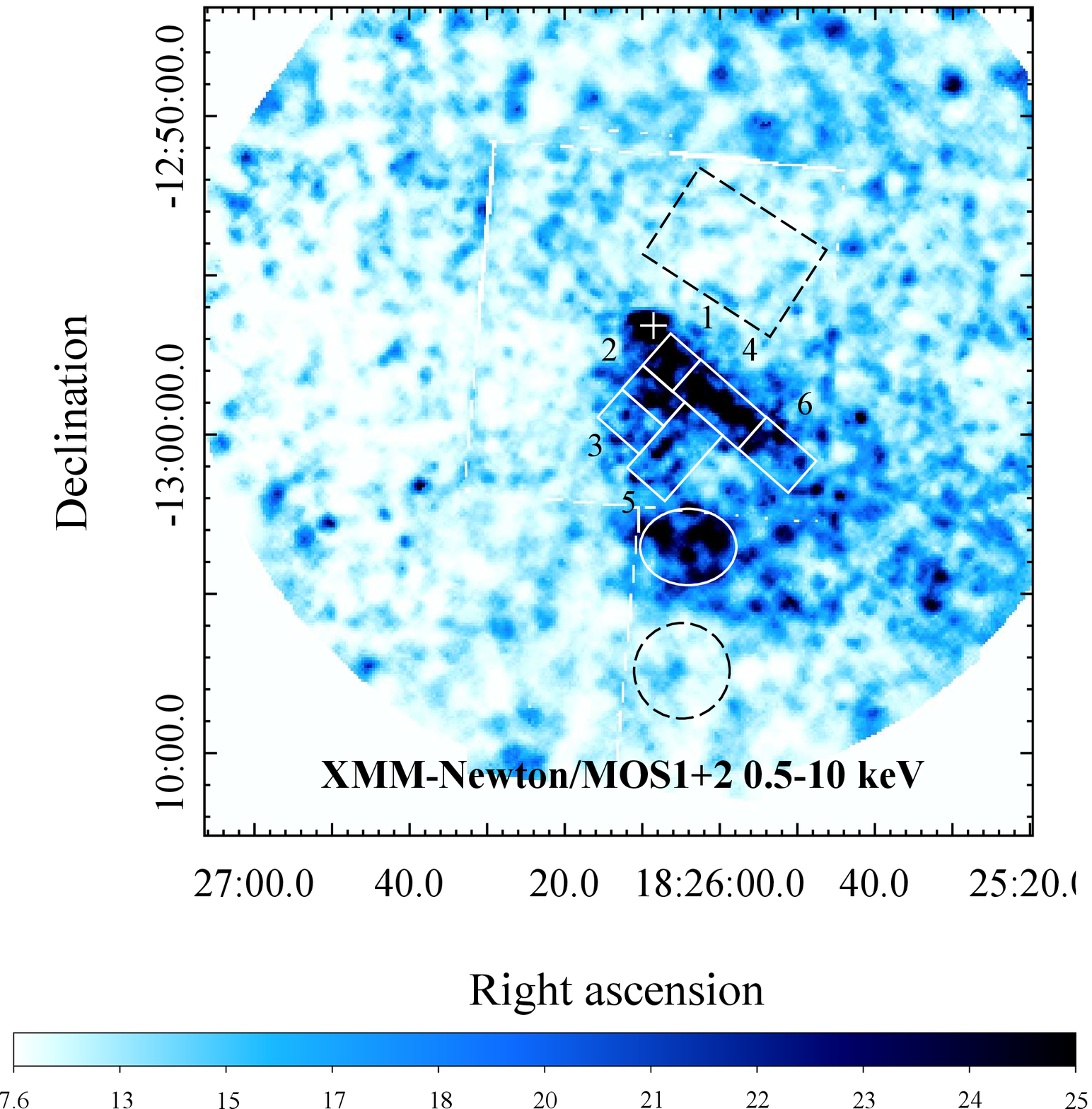}}
\end{minipage}
\caption{0.5--10 keV images of the \psr\ field. The pulsar \chan\ position is shown by the `+' symbol.
\textit{Top left}: Exposure-corrected FoV of the \chan\ ACIS-I detector. The intensity is given in 10$^{-8}$ ph~cm$^{-2}$~s$^{-1}$~pixel$^{-1}$. 
The black dashed box shows the EPIC-pn FoV.
The open star cluster Bica~3 is marked.
\textit{Top right}: FoV of the \xmm\ EPIC-pn camera in Small Window mode.
The dashed circle shows the background region for the pulsar spectral analysis.
The intensity is given in counts pixel$^{-1}$.
\textit{Bottom}: Combined  \xmm\ MOS1 and MOS2 exposure-corrected QPB-subtracted images.
In the bottom left panel bold solid, bold dashed, thin dashed and thin solid circles indicate
the positions and extents of the SNR candidate G18.45$-$0.42, 
open cluster Bica~3, HESS J1826$-$130 and SNR G18.6$-$0.2, respectively 
\citep[HESS J1826$-$130 has the size of 0.15 deg 
which is 1$\sigma$ for a single-Gaussian spatial model;][]{hesscollaboration2018}. 
In the bottom right panel we removed point sources except \psr.
Rectangular  regions used for the PWN spectral analysis are shown and numbered 
in  order of their distance from the pulsar.
The dashed polygon was used for the background.
The ellipse and circle show regions used to extract
spectra of Bica~3 and corresponding background, respectively.
The intensity is given in counts s$^{-1}$ deg$^{-2}$.
}
\label{fig:j1826comb}
\end{figure*}


\section{X-ray data and imaging}
\label{sec:data} 

A 140 ks \xmm\ observation of \psr\ field was carried out on 2014 October 
11 (ObsID 0744420101, PI Razzano).
European Photon Imaging Camera Metal Oxide Semiconductor
(EPIC-MOS) data were obtained in the Full Frame mode using the medium optical filter,
while the EPIC-pn camera was operated in the Small Window mode with the thin optical
filter\footnote{\url{ https://www.cosmos.esa.int/web/xmm-newton/technical-details-epic}}.
We also used \chan\ Advanced CCD Imaging Spectrometer (ACIS-I) observations performed on 
2003 February 17 (ObsID 3851, PI Romani, exposure time 15.1 ks, off-axis observation) 
and on 2007 July 26 (ObsID 7641, PI Roberts, exposure time 74.8 ks). 
We used the {\sc xmm-sas} v.16.0.0 and {\sc ciao} v.4.9 tools to reduce the data.

The \chan\ data were reprocessed using the {\sc chandra\_repro} tool.
In the top left panel of Fig.~\ref{fig:j1826comb} we show the field of view (FoV) 
of the ACIS-I\footnote{\url{http://asc.harvard.edu/proposer/POG/html/chap6.html}} detector 
in the 0.5--10 keV band created by the {\sc ciao fluximage} tool
from the longest \chan\ dataset (ObsID 7641). 
The `+' symbol marks the pulsar position 
(RA = 18\h26\m08\fs56, Dec = $-$12\degs56\amin34\farcs9) 
obtained using the {\sc wavdetect} tool.
The Eel trail extending south-west from \psr\ and  a fainter diffuse 
emission from the rest south part 
of the PWN are seen, along with brighter emission from the open star cluster Bica~3.

The \xmm\ data were reduced by the {\sc emchain} and {\sc epchain} tools
and periods of background flares were excluded by the {\sc espfilt} task with default parameters for the pn data and the {\sc mos-filter} command for the MOS data (the latter also runs {\sc espfilt}). 
The resulting exposures are 77.5, 81.2 and 49.8 ks for MOS1, MOS2 and pn, respectively.
The EPIC-pn FoV in the 0.5--10 keV band is shown 
in the top right panel of Fig.~\ref{fig:j1826comb}.
Since this camera was operated in the Small Window mode, it does not cover the whole PWN.
For comparison, its FoV is shown in the \chan\ image.

To obtain a wider and deeper image of the pulsar field, we combined data from both MOS detectors.
To do this, we utilize the \xmm\ Extended Source Analysis Software 
\citep[{\sc esas};][]{cookbook}.
The MOS images and exposure maps were generated by {\sc mos-spectra} tool 
(MOS1 CCDs 3 and 6 are no longer used,   
since they have been damaged by micrometeorites).
Quiescent particle background (QPB) images were created by the {\sc mos\_back} task.
Then MOS QPB-background subtracted and exposure corrected images were combined 
and adaptively smoothed by the {\sc adapt} tool
using 150 counts for the smoothing kernel. 
The resulting image in the 0.5--10 keV band is shown 
in the bottom left panel of Fig.~\ref{fig:j1826comb}. 
This is the deepest up-to-date soft X-ray image of the pulsar field.
The PWN emission, as well as the open star cluster Bica~3, 
is more visible than in the ACIS image.
Positions and extents of SNR candidate G18.45$-$0.42, 
HESS J1826$-$130 and SNR G18.6$-$0.2 are shown by circles. 
The former two  strongly overlap with each other and with the X-ray PWN, indicating their possible 
association, as has been mentioned in Section~\ref{intro},  
while the latter one has a large angular distance offset, suggesting that it is an unrelated object. 

To reveal the nebula emission better, we removed point-like sources
detected by the {\sc cheese} task in the \xmm\ data and 
by the {\sc wavdetect} command in the \chan\ datasets.
We refilled the resulting holes with values of the surrounding background emission
using the {\sc ciao dmfilth} tool.
The  exclusion of \psr\ distorts the shape of the PWN so it was not removed.
The resulting image smoothed using the smoothing kernel of 150 counts
is presented in the bottom right panel of Fig.~\ref{fig:j1826comb}.  
In the \xmm\ data, the Eel trail is detected up to 6 arcmin from the pulsar, 
while in the less-sensitive \chan\ observations only up to 4 arcmin.
Fainter diffusive nebula emission  in the south  may be somewhat blended with Bica~3. 
Overall,  the nebula resembles a cometary-like tail behind \psr\ which is 
typical for PWNe created by fast-moving pulsars \citep{kargaltsev2017}.


\section{Spectral analysis}
\label{sec:spectra}

To perform pulsar spectral analysis,
we extracted spectra from MOS and pn data using
the {\sc evselect} tool and the 15-arcsecs radius circle around the \psr\ \chan\ 
position\footnote{The aperture size was selected using the {\sc eregionanalyse} tool.}.
For the background, we used the region free from any sources  
(top right panel of Fig.~\ref{fig:j1826comb}).
The {\sc rmfgen} and {\sc arfgen} tasks were applied to generate
the redistribution matrix and ancillary response files.
The spectra were grouped to ensure 20 counts per energy bin.
We also extracted the \psr\ spectrum from the longest \chan/ACIS-I dataset
utilizing the {\sc ciao specextarct} tool and the 2-arcsecs radius 
circle (the short off-axis \chan~observation provides much smaller number of counts and was neglected).
It was also grouped to ensure 20 counts per energy bin.
We obtained 352(MOS1)+383(MOS2)+739(pn)+207(ACIS) source counts in the 0.5--10 keV band.
The \xmm\ spectra contains both the pulsar and the adjacent PWN emission,
due to the broad wings of the \xmm\ point spread function (PSF).
Thus, we generated the response files for a point source 
as well as for an extended source.
To constrain the PWN contribution in the \xmm\ spectra,
we extracted its spectrum from the \chan\ data using
an annulus with inner and outer radii of 2.5 and 15 arcsec.
This resulted in 206 source counts 
which were grouped to ensure 20 counts per energy bin.
To analyse  the \psr\ spectra, we tried two different models:
a power-law (PL) model which describes non-thermal magnetosphere 
emission and a black body (BB) model which describes 
thermal emission from the pulsar surface.
The pulsar model was convolved with the response files for a point source.
The PWN contribution was described by the PL model 
convolved with the response files for an extended source.
We added a constant factor in the models, which represents 
the relative normalization between \chan\ and \xmm\ detectors.

To perform spatial-resolved spectral analysis of the Eel nebula at a larger scale,
we extracted source and background spectra from the MOS data using regions
shown and numbered in the right panel of 
Fig.~\ref{fig:j1826comb}\footnote{
We studied that part of the PWN which is covered by both MOS detectors
(in the south-west it is partially projected onto a disabled MOS1 CCD).
We did not use \chan\ data for the nebula spectral analysis
due to the gaps between ACIS-I CCDs.}.
Point sources were excluded from these regions. 
The spectra of the  regions were grouped to ensure 30 counts per
energy bin
and were analysed using  PL models.

\begin{figure}
\begin{minipage}[h]{1.\linewidth}
\center{\includegraphics[width=0.69\linewidth,angle=-90,clip]{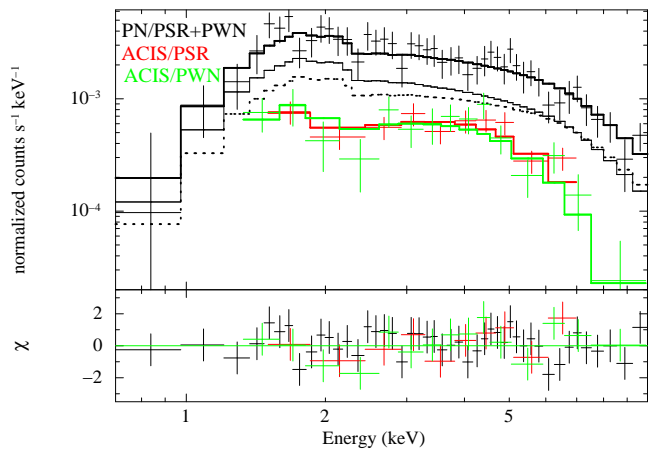}}
\end{minipage}
\caption{\textit{Top}: Observed X-ray spectra of \psr\ and  its PWN 
within 15 arcsec from the pulsar.
The data from different instruments and sources are shown by different colours. 
Bold solid lines show the best-fit models. 
Dotted and thin solid lines show the pulsar and PWN contribution in the \xmm\ pn-spectrum, respectively.
\textit{Bottom}: Fit residuals.}
\label{fig:psrspec}
\end{figure}

\begin{figure}
\begin{minipage}[h]{1.\linewidth}
\center{\includegraphics[width=1.\linewidth,clip]{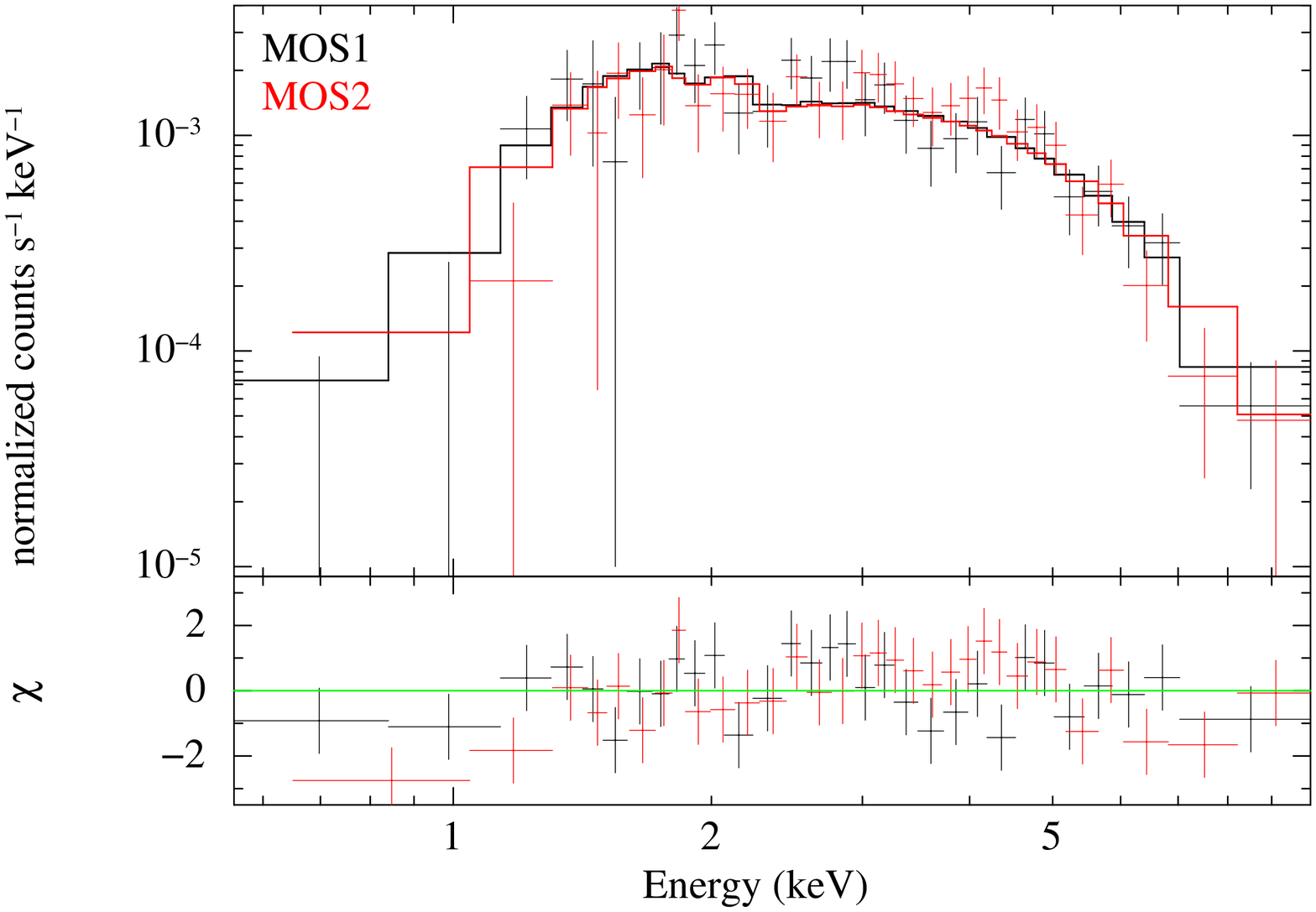}}
\end{minipage}
\begin{minipage}[h]{1.\linewidth}
\center{\includegraphics[width=1.\linewidth,clip]{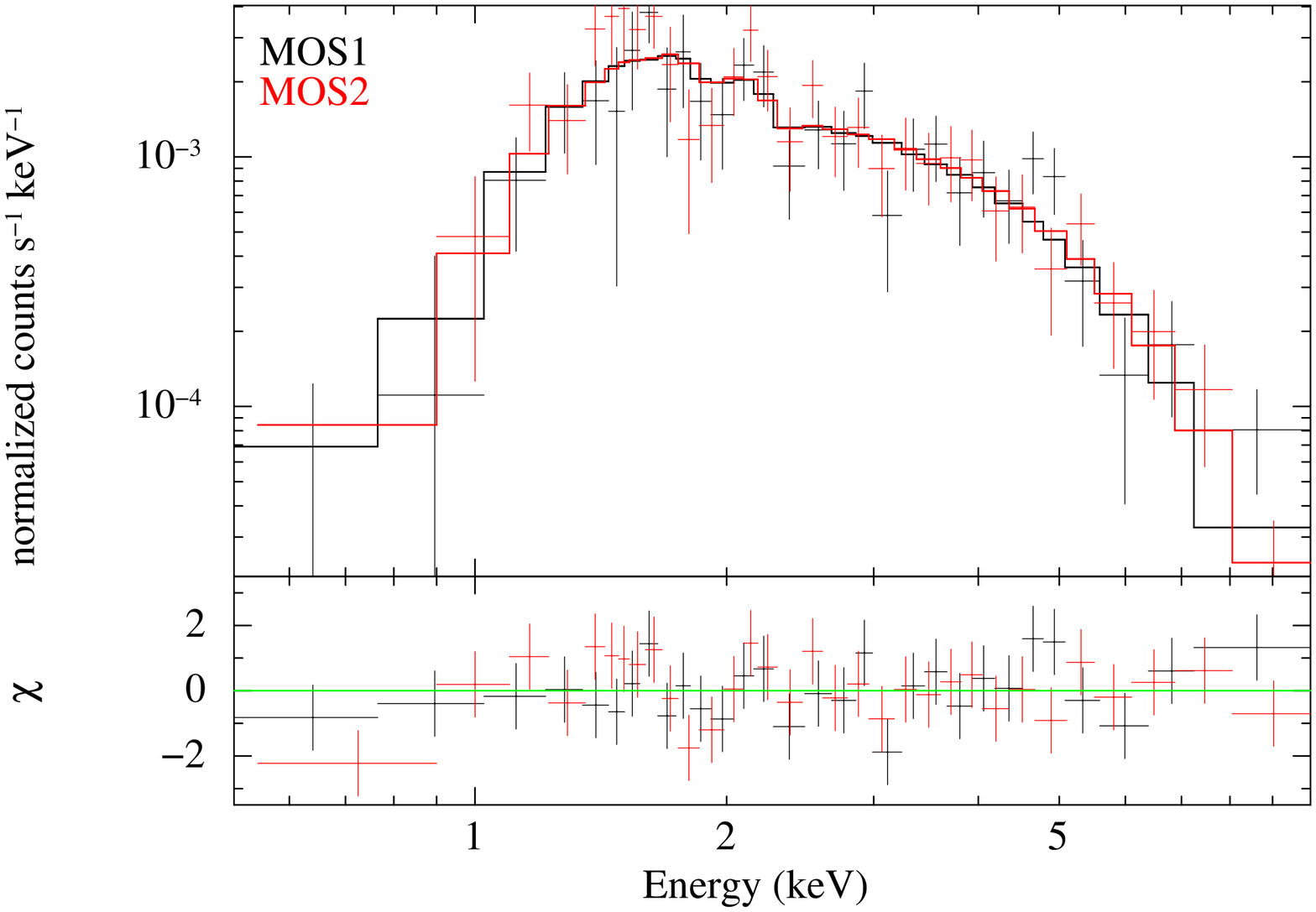}}
\end{minipage}
\caption{Examples of observed X-ray spectra 
of the Eel nebula extracted from regions 1 (top) and 4 (bottom).
The data from different instruments are shown by different colours. 
Solid lines show the best-fit PL models.}
\label{fig:pwnspec}
\end{figure}

All spectra (including the pulsar, the adjacent  PWN,  
i.e. within 15 arcsec from the pulsar and its more distant regions) 
were fitted simultaneously in the 0.5--10 keV band using {\sc xspec} v.12.9.1 and assuming 
a common value of the absorption column density $N_{\rm H}$.
For photoelectric absorption, the {\sc xspec} Tuebingen-Boulder
interstellar medium (ISM) absorption model {\sc tbabs} with the {\sc wilm} 
abundances \citep{wilms2000} was used.
We found that both PL and BB models chosen for the pulsar 
can describe its spectra well. 
However, the BB model resulted in temperature $T\approx1.4$ keV 
which is too large for pulsar thermal emission 
either from the whole stellar surface or a  polar cap \citep[see e.g.][]{vigano2013}.
Thus, we rejected this model.
The resulting best-fitting parameters are presented in 
Tables~\ref{tab:psrpar} and \ref{tab:pwnpar}.
We obtained $\chi^2=511$ for 549 degrees of freedom (d.o.f.).
The spectra of the pulsar and the adjacent PWN are shown in Fig.~\ref{fig:psrspec}.
The examples of the PWN spectra from the  regions 1 and 4 are shown in Fig.~\ref{fig:pwnspec}.

We also tried to use the BB+PL model for the pulsar, but this did not lead 
to substantial improvements of the fit statistics:    
{\sc ftest} command provided a probability of about 0.4 that the improvements occurred by chance.

\begin{table}
\renewcommand{\arraystretch}{1.2}
\caption{Best-fitting parameters for \psr\ and adjacent PWN (i.e. within 15 arcsec from the pulsar) emission.
All errors are at 90\% confidence.
Unabsorbed fluxes $F_X$ in the 0.5--10 keV band were calculated using {\sc xspec} convolution model {\sc cflux}.
$N_{\rm bins}$ = number of spectral bins. 
The cross-normalization constant was fixed at 1 for \xmm\ spectra
and free for \chan\ spectra.}
\label{tab:psrpar}
\begin{center}
\begin{tabular}{lc}                         
\hline 
Column density $N_{\rm H}$, $10^{22}$ cm$^{-2}$                & $2.2^{+0.2}_{-0.2}$ \\
\hline 
Pulsar photon index $\Gamma^{\rm psr}$                         & $0.92^{+0.25}_{-0.24}$ \\
Pulsar flux $F_X^{\rm psr}$, $10^{-13}$ erg s$^{-1}$ cm$^{-2}$ & $1.04^{+0.14}_{-0.13}$ \\
\hline 
PWN photon index $\Gamma^{\rm pwn}$                            & $1.20^{+0.24}_{-0.23}$ \\
PWN flux $F_X^{\rm pwn}$, $10^{-13}$ erg s$^{-1}$ cm$^{-2}$    & $0.85^{+0.10}_{-0.09}$ \\
\hline 
Cross-normalization constant                                   & $1.14\pm0.13$ \\
$\chi^2$($N_{\rm bins}$)                                       & 94(114)  \\
\hline
\end{tabular}
\end{center}
\end{table}

\begin{table*}
\renewcommand{\arraystretch}{1.2}
\caption{Best-fitting parameters for the Eel nebula emission (regions 1--6 in Fig.~\ref{fig:j1826comb}).
All errors are at 90\% confidence. 
Unabsorbed fluxes $F_X$ are given in the 0.5--10 keV band.
$N_{\rm bins}$ = number of spectral bins.}
\label{tab:pwnpar}
\begin{center}
\begin{tabular}{lcccccc}                         
\hline 
Region           							  & 1                      & 2                      & 3                      & 4                      & 5                      & 6 \\
\hline
Net counts (MOS1/2)                                                       & 500/513                & 435/476                & 185/239                & 953/1036               & 464/479                & 337/369 \\
\hline
Photon index $\Gamma$                                                     & $1.34^{+0.18}_{-0.17}$ & $1.99^{+0.23}_{-0.22}$ & $2.44^{+0.41}_{-0.37}$ & $1.76^{+0.16}_{-0.16}$ & $2.54^{+0.33}_{-0.31}$ & $2.58^{+0.36}_{-0.32}$ \\
PL normalization $K_{\rm PL}$, $10^{-5}$ ph s$^{-1}$ cm$^{-2}$ keV$^{-1}$ & $2.2^{+0.6}_{-0.5}$    & $3.9^{+1.2}_{-0.9}$    & $2.7^{+1.2}_{-0.9}$    & $6.7^{+1.5}_{-1.2}$    & $2.7^{+1.2}_{-0.9}$    & $5.5^{+2.2}_{-1.5}$    \\
Flux $F_X$, $10^{-13}$ erg s$^{-1}$ cm$^{-2}$                             & $2.07^{+0.16}_{-0.17}$ & $1.87^{+0.24}_{-0.19}$ & $0.96^{+0.27}_{-0.18}$ & $3.99^{+0.34}_{-0.29}$ & $2.30^{+0.58}_{-0.38}$ & $1.88^{+0.50}_{-0.33}$ \\
Area, arcmin$^{2}$                                                        & 1.4                    & 1.5                    & 1.3                    & 3.3                    & 3.0                    & 2.2 \\
Distance from the pulsar, arcmin                                          & 1.3                    & 2.2                    & 3.1                    & 3.2                    & 4.0                    & 5.6 \\
\hline 
$\chi^2$($N_{\rm bins}$)                                                  & 60(63)                 & 46(62)                 & 39(40)                 & 116(128)               & 81(92)                 & 74(68)\\
\hline
\end{tabular}
\end{center}
\end{table*}

\section{Discussion}
\label{sec:discussion}

\subsection{Distance to \psr}
\label{subsec:dist}

To constrain the parameters of the pulsar and its PWN,
it is necessary to know the distance.

The only available `pseudo'-distance estimate $D\sim 1.2$~kpc \citep{marelliphd} 
is known to be uncertain within a factor of 2--3.
Assuming a 100 per cent efficiency in $\gamma$-rays, we obtain $D\approx9.4$ kpc.
However, this value can be even larger, accounting for an  unknown emission beaming factor. 
Using the pulsar galactic coordinates $(l,b)=(18.56,-0.38)$ and  
assuming that it lies inside the Galactic disk with a half-thickness of 100 pc,
the maximum $D$ is $\sim 15$ kpc if the pulsar is located near the disk  edge.

A more reliable distance estimate can be obtained by using the empirical relation 
between the distance and the interstellar reddening $E(B-V)$ in the pulsar direction.
According to the extinction map by 
\citet{schlafly2011}\footnote{\url{https://irsa.ipac.caltech.edu/applications/DUST/}},
the total Galactic $E(B-V)$ in this direction is about 15.5.
The reddening to \psr\ can be derived from the column density $N_{\rm H}$ value utilizing the relation 
$N_{\rm H}=(8.9\pm 0.4)\times E(B-V)\times 10^{21}$~cm$^{-2}$ \citep{foight2016}.
$N_{\rm H}$  obtained from the X-ray spectral analysis (Table~\ref{tab:psrpar})
transforms to $E(B-V)=2.45\pm0.19$ (1$\sigma$ errors). This is much smaller than the total Galactic 
value and implies that the pulsar is much closer than the Galactic disk edge in its direction. 
We compared the obtained $E(B-V)$ with the extinction map by \citet{Marshall2006},
which is based on Two-Micron All-Sky Survey (2MASS) stars photometry along 
with the Besan\c{c}on model of population synthesis. 
The $E(B-V)$--distance relation was constructed 
using the {\sc python} package {\sc mwdust} \citep{Bovy2016}.
The resulting  $D$ lies in the range of 3.4--3.6 kpc. 
This is compatible with the distance estimate of 3 kpc based on the pulsar  
X-ray luminosity, though such estimate is very uncertain \citep{roberts2009}. 
We adopt 3.5 kpc as a reasonable value in  our following estimates. 


\subsection{Pulsar}
\label{subsec:psr}

The \psr\ spectrum can be described by the PL model.
The obtained photon index and flux (Table~\ref{tab:psrpar}) are in agreement 
with the results of \citet{marelliphd}, who used only \chan\ data.
Addition of the \xmm\ data allowed us to constrain these parameters better.
They are different from the results by \citet{Li2018apjl} 
since they used only the MOS data and did not take into account 
the PWN contribution in the pulsar aperture. 

The pulsar non-thermal X-ray luminosity $L_{X}=(1.5\pm 0.2)\times10^{32}D^{2}_{3.5}$ erg~s$^{-1}$ 
and efficiency $\eta_{X}=L_{ X}/\dot{E}=(4.2\pm0.5)\times10^{-5}D^{2}_{3.5}$ in the 0.5--10 keV band,
where $D_{3.5}$ is the distance in the units of 3.5 kpc.
The values obtained are typical for pulsars with similar ages and spin-down luminosities, 
although they are by an order of magnitude higher than those of the Vela pulsar, 
which is known as a very inefficient non-thermal emitter \citep[e.g.,][]{kargaltsev2008}.
The ratio between the $\gamma$-ray \citep{marelliphd} 
and the non-thermal unabsorbed X-ray fluxes 
of the pulsar is ${\rm log}(F_\gamma/F_X)\approx3.5$.
This is in agreement with the value of 3.5$\pm$0.5 derived for  radio-quiet
$\gamma$-ray pulsars \citep{abdo2013, marelli2015}.

We estimated the \psr\ surface temperature adding the BB component
to the pulsar non-thermal spectral model and adopting a neutron star (NS) radius of 13$D_{3.5}$ km. 
We obtained a 3$\sigma$ upper limit on the temperature of $\approx0.1$ keV, which
is consistent with predictions of standard NS cooling scenarios    
for a 14 kyr star \citep{yakovlev2004}. 


\subsection{PWN}
\label{subsec:pwn}

The spectrum of the PWN within $\approx$ 15 arcsec from the pulsar, 
with the photon index $\Gamma$ = 1.2$\pm$0.2 (see Table~\ref{tab:psrpar}), is hard. The same is observed
in other PWNe powered by Vela-like pulsars \citep{Bykov2017SSRv}. 
The index increases to $\approx$2.5 with the distance from the pulsar (Table \ref{tab:pwnpar}).
Such spectral steepening is observed 
for some other PWNe and suggests synchrotron 
cooling \citep[e.g.][]{reynolds2017,slane2017hb}.
The total PWN luminosity is about $L_{X} \approx 2\times10^{33}$
$D_{3.5}^{2}$ erg~s$^{-1}$. This corresponds
to the X-ray efficiency of  $5.6\times10^{-4}D_{3.5}^{2}$, which is typical for PWNe. 

\begin{figure}
\begin{minipage}[h]{1.\linewidth}
\center{\includegraphics[width=1.\linewidth, clip]{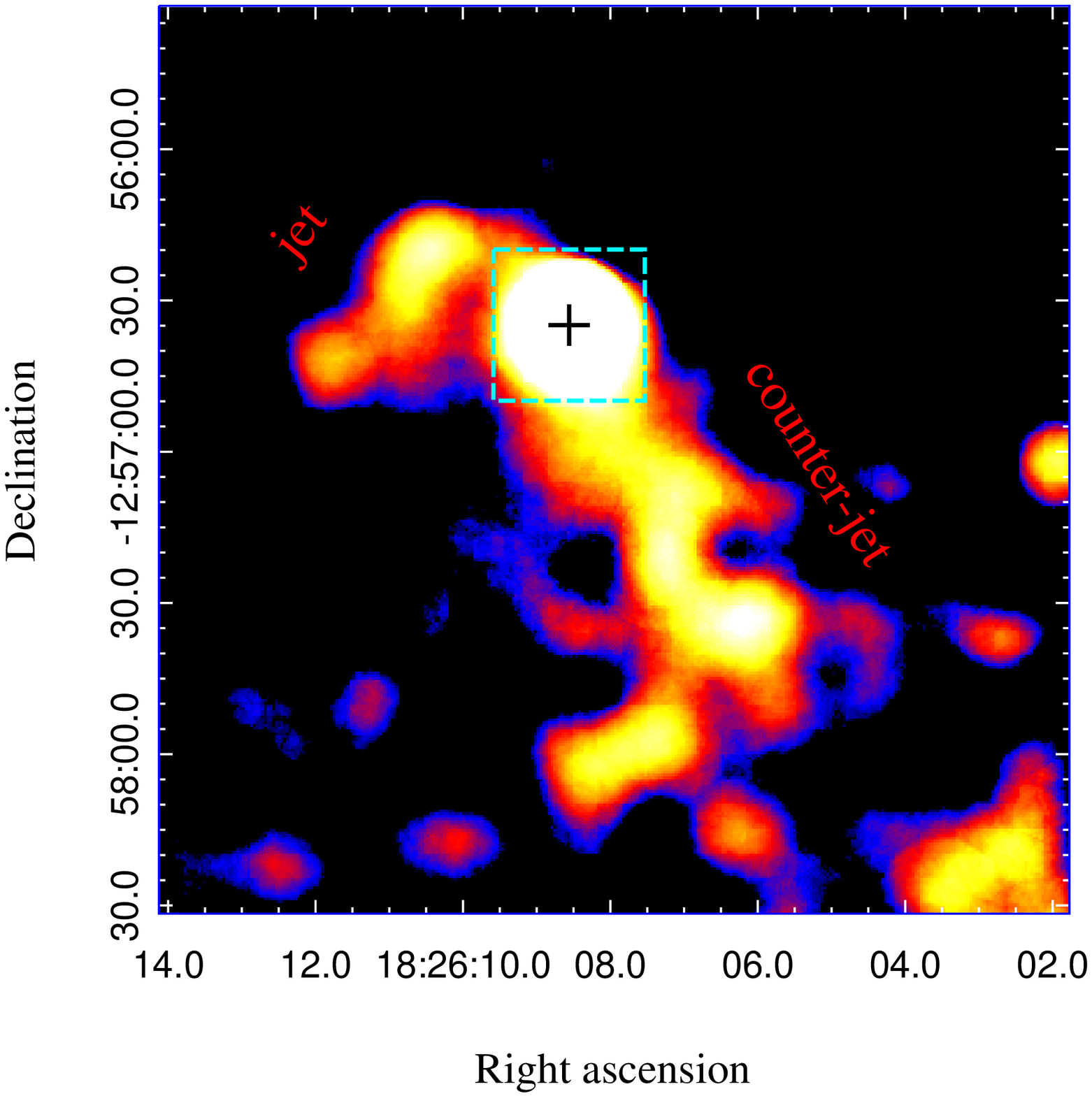}}
\put (-200,40) {\includegraphics[scale=0.15, clip]{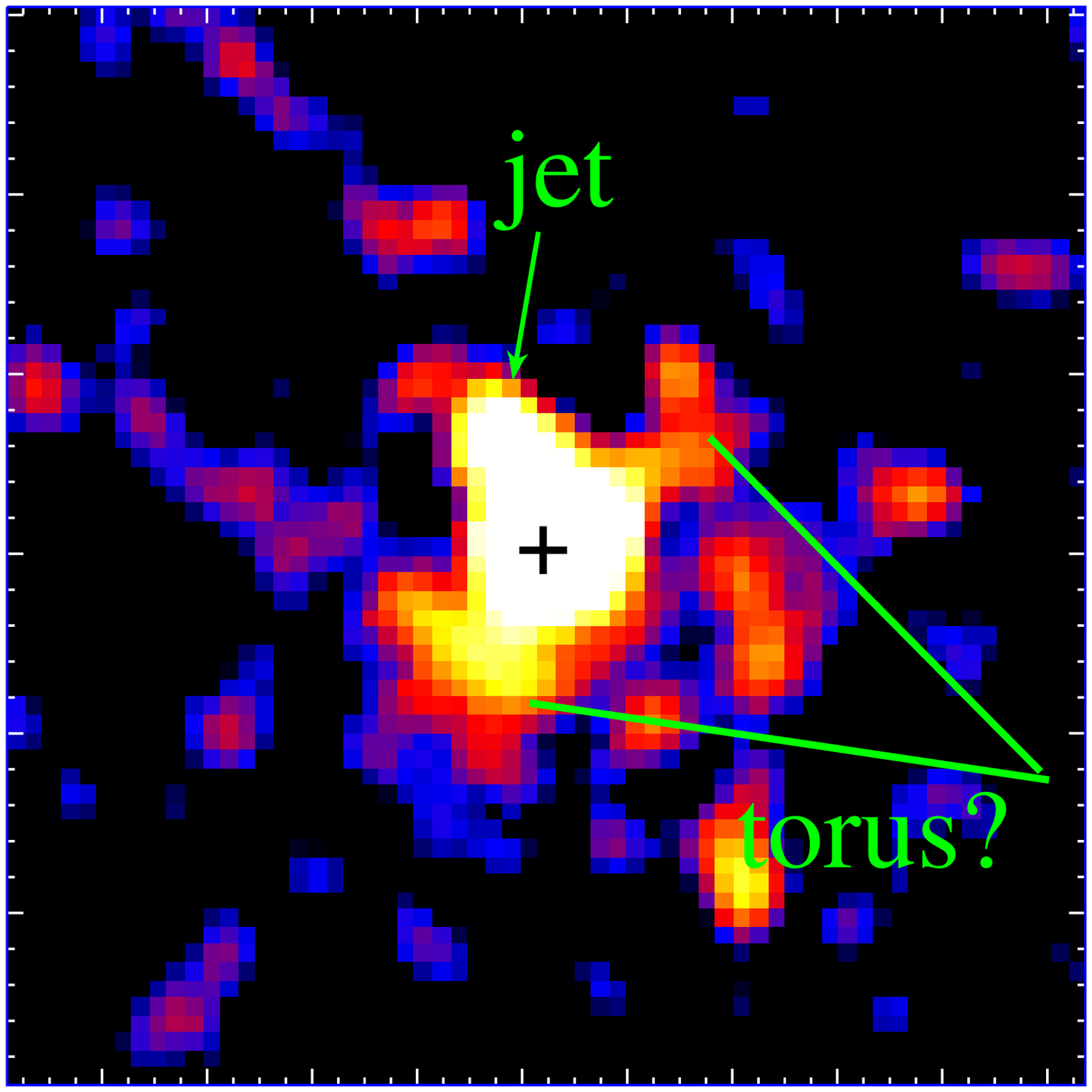}}
\end{minipage}
\caption{3 arcmin $\times$ 3 arcmin 
\chan\ X-ray image of the pulsar vicinity in 0.5--10 keV range smoothed with 
a 25 pixel Gaussian kernel.
The `+' symbol shows the pulsar position.
The `jet' and `counter-jet'
are marked. 
The 30 arcsec $\times$ 30 arcsec image part, enclosed by the cyan dashed box and
smoothed with a 3 pixel Gaussian kernel, is enlarged in the inset. 
The possible PWN torus and
the base part of the `jet' are marked. 
}
\label{fig:torus}
\end{figure}

\begin{figure}
\begin{minipage}[h]{1.\linewidth}
\center{\includegraphics[width=0.87\linewidth, clip]{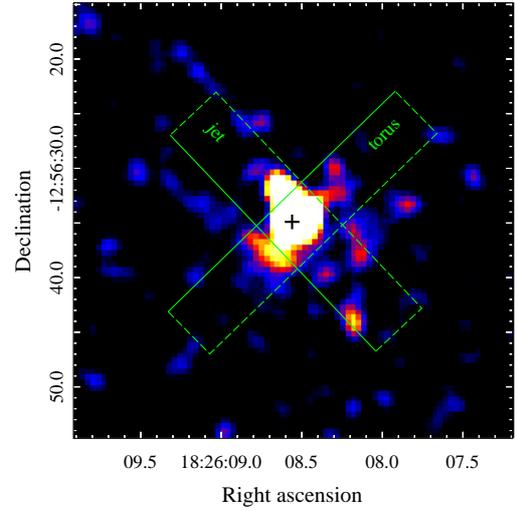}}
\end{minipage}
\begin{minipage}[h]{1.\linewidth}
\center{\includegraphics[width=0.83\linewidth, clip]{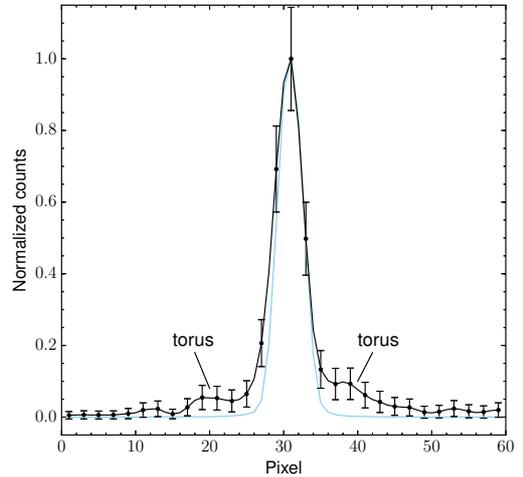}}
\end{minipage}
\begin{minipage}[h]{1.\linewidth}
\center{\includegraphics[width=0.83\linewidth, clip]{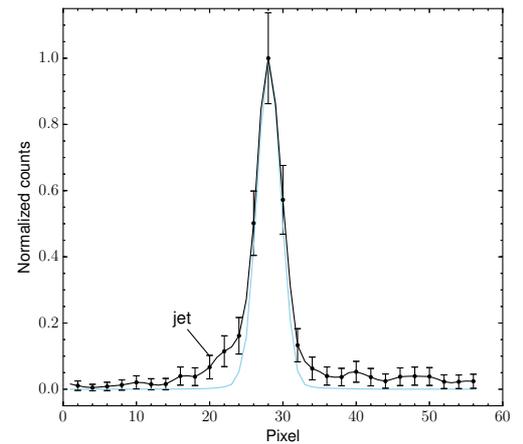}}
\end{minipage}
\caption{\chan\ X-ray image of the pulsar vicinity in 0.5--10 keV range 
smoothed with a 3 pixel Gaussian kernel (top).
The `+' symbol shows the pulsar position.
Green rectangles are used to obtain spatial profiles along the presumed torus and jet structures shown in the middle and bottom panels, respectively. Points (0,0) in horizontal  axes correspond to  the top-short  rectangle sides. The data are shown in black and the PSF -- in light-blue. 
The background level is negligible.
}
\label{fig:proj}
\end{figure}


\begin{figure}
\begin{minipage}[h]{1.\linewidth}
\center{\includegraphics[width=1.\linewidth,clip]{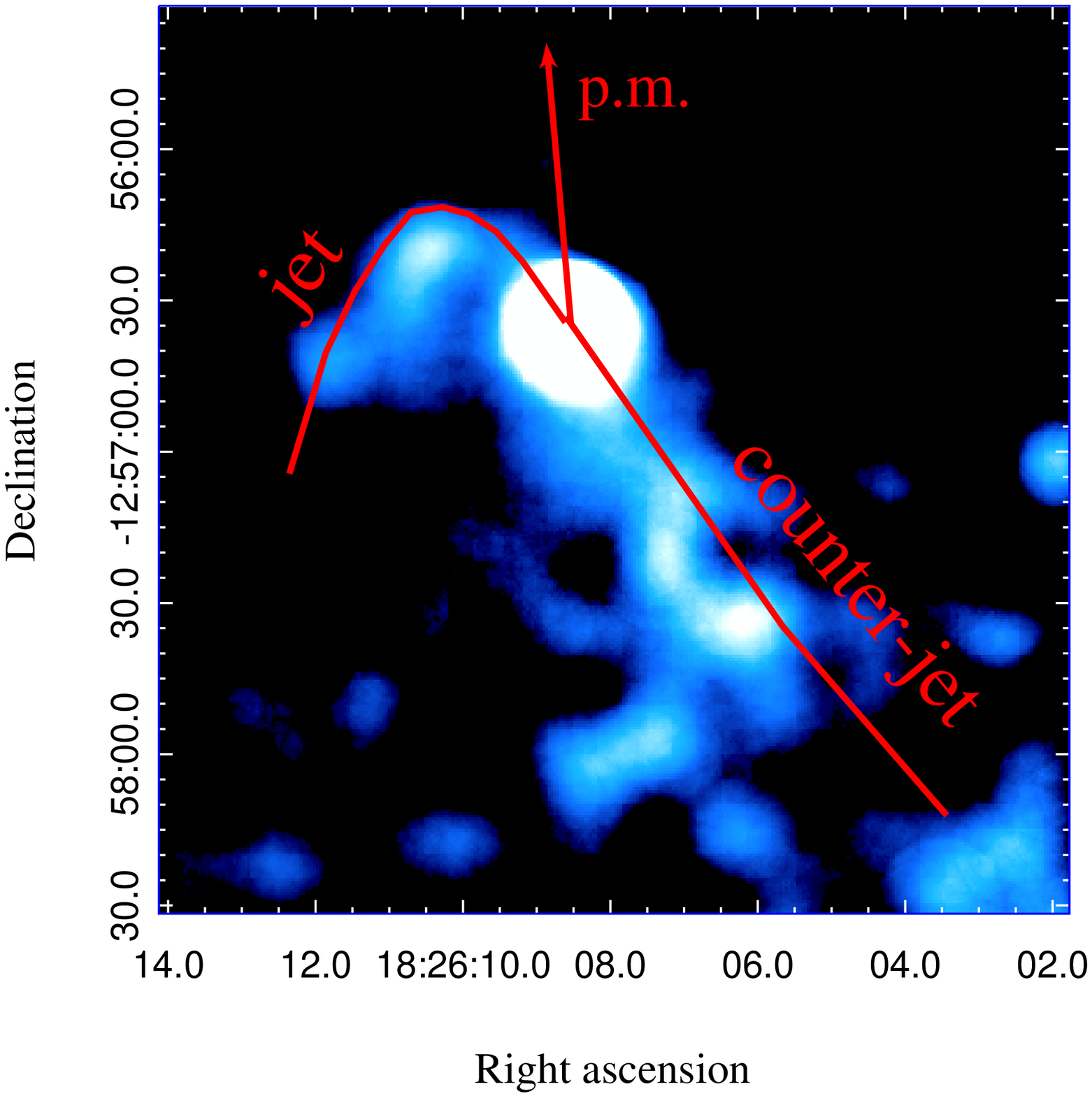}}
\end{minipage}
\begin{minipage}[h]{1.\linewidth}
\center{\includegraphics[width=1.\linewidth,clip]{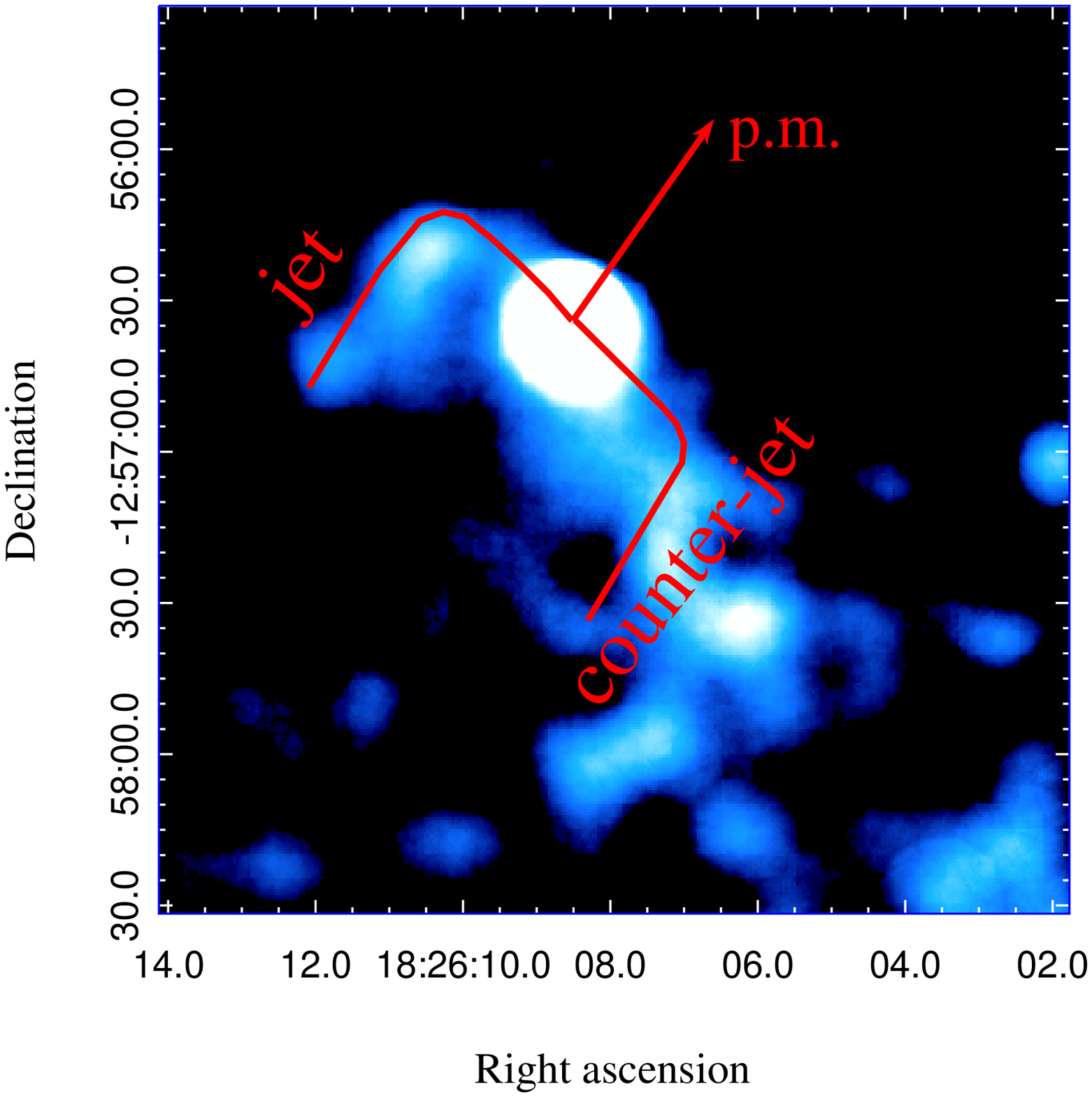}}
\end{minipage}
\caption{The same image as in Fig. \ref{fig:torus} but using another color map.
Possible geometries of the system are shown by the red lines 
as suggested in this work (top) and in \citet{kargaltsev2017} (bottom).
Arrows shows the suggested pulsar \pmo\ (p.m.) directions. 
}
\label{fig:2x2}
\end{figure}

\citet{roberts2009} noticed X-ray structures in the pulsar's near vicinity,
which can be interpreted as the PWN torus and jet.
By considering the \chan\ data, we confirm the presence of the presumed torus around the pulsar, 
with the radius of about 5 arcsec, likely seen edge on, and the north-east jet 
directed along the torus axis (the insert of Fig.~\ref{fig:torus}).
In Fig.~\ref{fig:proj} we show spatial profiles along these structures,
compared with the point spread function (PSF). 
The latter was generated using {\sc chart} \citep{chart} and {\sc marx} tools 
and the \psr\ best-fitting spectral model 
(Table~\ref{tab:psrpar})\footnote{For details see \url{http://cxc.harvard.edu/ciao/PSFs/chart2/index.html}}. 
Emission excesses over the PSF wings corresponding to the torus and 
jet structures adjacent to the PSF core are clearly seen.
Such structures are observed in other young PWNe, including the Vela PWN.
In the latter case, the size of the torus  is about 2 arcmin \citep{Velasize}.
At a distance of 3.5 kpc, this would transform to $\approx10$ arcsec,
which is compatible with the observed dimension of the presumed \psr\ torus,
suggesting that it is a real PWN structure.
Assuming that, we can estimate the upper limit of  
the radius of the pulsar wind termination shock (TS) which should be smaller  
than the torus:  $R_{\rm TS} \lesssim 2.6\times 10^{17} D_{3.5}$ cm.  
This is comparable with that of the Vela PWN \citep[$5\times10^{16}$ cm;][]{kargaltsev2008}.
The respective lower limit on the ambient  matter pressure is  
P$_{\rm amb}$ $\approx \dot{E}(4 \pi c R^{2}_{\rm TS})^{-1}$
$\approx$ $0.14\times 10^{-9}(R_{\rm TS}/2.6\times10^{17}~{\rm cm})^{-2}$ dyn~cm$^{-2}$.
This value is typical for Vela-like pulsars located inside their host SNRs \citep{2007ApJ...660.1413K}.
For \psr, this host could be the SNR candidate G18.45$-$0.42 (see details below).

At a larger scale, we see that the jet mentioned above is bent towards the east, 
representing the `Eel head' (the main panel of Fig.~\ref{fig:torus}).
\citet{roberts2009} referred it as a `forward jet'.  
The longer `Eel body' extending about 6 arcmin ($\approx$ 6D$_{3.5}$ pc) 
south-west of the pulsar can be 
interpreted as a `counter-jet', which transforms at larger distances 
to a trail behind the pulsar moving in roughly the opposite direction.
The corresponding geometry is shown in the top panel of Fig.~\ref{fig:2x2}.
The bending of the forward jet could be due to ram pressure if   
the vector of the pulsar velocity directs approximately to the north. 
A similar interpretation is considered for  
the Vela PWN, where the pulsar \pmo\ and the matter 
flow initiated by the SNR reverse shock are likely working 
together to bend the PWN jets \citep{Kargaltsev2015SSRv}. 
To conclude, the Eel appears to be a mixed type morphology PWN
containing a torus, jets and a trail.

Note that \citet{kargaltsev2017} suggested another geometry of the system, which is
shown in the bottom panel of Fig.~\ref{fig:2x2} (see also their fig.~6).
The corresponding \pmo\ position angle (P.A.) is $\sim$330 deg.
However, this geometry cannot describe the long PWN trail.
Moreover, only the jet is clearly bent while the counter-jet is not.


\subsection{Connection between \psr\ PWN, the SNR candidate G18.45$-$0.42, 
the open cluster Bica~3 and HESS J1826$-$130}
\label{subsec:psr-snr}

\psr\ and the Eel nebula are located in the complex region
containing the SNR candidate G18.45$-$0.42, the TeV source HESS J1826$-$130 and
the open stellar cluster Bica~3 (Fig.~\ref{fig:j1826comb}).
Below we discuss possible associations between them.

It was proposed earlier that HESS J1826$-$130 
is a TeV counterpart to the \psr\ PWN
\citep{roberts2007,anguner2017o}.
This suggestion is natural, since there are similar Vela-like systems.
One remarkable example is PSR J1823$-$13 
\citep*[$\dot{E}=2.8\times10^{36}$ erg~s$^{-1}$, $\tau=21$ kyr;][]{pavlov2008}.
Measurements of this pulsar's \pmo\ allowed authors to understand 
its connection with HESS J1825$-$137.  
\citet{pavlov2008} suggested that the latter is produced by relic electrons
emitted by the pulsar when it was younger and more powerful.
Another example is the young and energetic pulsar J1357$-$6429 
($\dot{E}=3.1\times10^{36}$ erg~s$^{-1}$, $\tau=7.3$ kyr).
The pulsar and its PWN are projected onto the extended TeV source HESS J1356$-$645
\citep{abramowski2011,chang2012}.
\citet{chang2012} proposed the most plausible explanation 
of the TeV and radio emission as arising from a relic PWN.

The Eel-TeV-source association is also supported by the fact 
that the PWN X-ray and TeV luminosities, $L_X\approx 2\times10^{33}$
$D_{3.5}^{2}$~erg~s$^{-1}$ (this paper) and $L_\gamma\approx 1.5\times10^{33}$
$D_{3.5}^{2}$~erg~s$^{-1}$ \citep{hesscollaboration2018},     
are consistent with the $L_{\gamma \rm >TeV}$ versus $L_X$ and $L_{\gamma \rm >TeV}/L_X$ 
versus age distributions of other X-$\gamma$-ray PWNe  \citep{kargaltsev2013}.
In addition, if we assume that the TeV source is the relic PWN of J1826 then Bohm diffusion 
can be the main process of the particle escape and the source expansion.
On the time-scale of 14 kyr the particles will diffuse to a distance of 
$40B_{-6}^{-1/2} (1+0.144 B^{2}_{-6})^{-1/2}$ pc,
where $B_{-6}$ is the magnetic field in $\mu$G \citep{kargaltsev2013}. 
For the visible TeV source size of about $9 D_{3.5}$ pc, this translates to a magnetic field
of about 5 $\mu$G, which is a typical value for PWNe.

There is some displacement between the TeV source centres  
and the pulsars in the systems noticed above,
which is usually explained by the effect of the SNR reverse shock 
on the PWN and/or by the pulsar \pmo\ \citep{blondin2001}. 
The same situation is observed for \psr\ and HESS J1826$-$130 (Fig.~\ref{fig:j1826comb}).
Their displacement could be explained naturally by assuming that 
the SNR candidate G18.45$-$0.42 is the pulsar host remnant.  
\psr\ is located about 7.5 arcmin off the remnant centre
and projects onto its shell (Fig.~\ref{fig:j1826comb}).
If the pulsar was born near the  G18.45$-$0.42 centre, 
for an age of $\approx 14$~kyr    
its \pmo\ has to be $\approx 32t^{-1}_{14}$ mas~yr$^{-1}$,
implying the pulsar transverse velocity of $\approx 530D_{3.5}$ km~s$^{-1}$. 
The latter value is consistent with the pulsar velocity distribution \citep{hobbs2005}.
The \pmo\ P.A. in this case is $\sim$5 deg, 
which is compatible with  the geometry explaining the Eel structure presented in the top panel 
of Fig.~\ref{fig:2x2}.
The alternative  geometry (the bottom panel of Fig.~\ref{fig:2x2}) considered by \citet{kargaltsev2017} raises  
a question about the \psr\ host SNR. In this case, 
the pulsar cannot be associated with any remnant found around it  
(Fig.~\ref{fig:j1826comb}).

If G18.45$-$0.42 is the real SNR related with the pulsar, its observed radius is  $\sim$ 8D$_{3.5}$ pc. 
At the age of about 14 kyr, it has to be entered into the pressure driven snow-plough phase. 
Using the SNR evolution code \citep{leahy2017} with typical ISM number 
density  of 0.5--2 cm$^{-3}$ and  supernova explosion energy  of $10^{51}$ erg 
we obtain the blast wave shock radius of about 12--14 pc. Accounting for uncertainties 
in the age and distance, this is compatible with the observed radius.      
Deeper study of  G18.45$-$0.42 is necessary 
to confirm its SNR nature and the association with \psr\ and to estimate its distance and age.   

We have tried to measure the pulsar \pmo\ 
using both \chan\ observations, providing a time base of 4.4 yr.
To perform the astrometric transformation, we used stars
detected in both datasets by the {\sc wavdetect} tool 
with significance $>9\sigma$ (sources detected at chips edges were excluded).
The short observation was registered to the long one as described in the \chan\ 
manual\footnote{\url{http://cxc.harvard.edu/ciao/threads/reproject_aspect/}}.
We obtained only an upper limit on the pulsar 
position shift of $\lesssim0.7$ arcsec corresponding  
to a non-informative \pmo\ limit of $\lesssim160$ mas~yr$^{-1}$ as compared to  
the expected value of $\approx32$ mas~yr$^{-1}$ estimated above. 
The are two reasons for such a rude result.
The first one is the short exposure of the \chan\ observation (ObsID 3851)
leading to a non-sufficient signal to noise ratio of $\approx 6$. 
The second is that in this observation the  pulsar was  exposed 
at $\approx5.5$ arcmin from the telescope aim-point, where 
the point source localization accuracy degrades significantly.
Additional \chan\ observations are necessary to measure the \pmo. 

The Bica~3 position coincides with the center of G18.45$-$0.42
which makes it a likely birthplace of this presumed SNR. 
According to the OPENCLUST catalogue \citep{dias2002}, 
the distance to the cluster is about 1.6 kpc.
We extracted the cluster spectrum from MOS2 data 
using the region shown in the bottom right 
panel of Fig.~\ref{fig:j1826comb} 
(in the case of MOS1, Bica~3 is projected on the disabled CCD).
It can be described by the absorbed model of 
collisionally-ionized diffuse gas {\sc apec}
with column density $N_{\rm H}=7.6^{+1.7}_{-1.4}\times10^{21}$ cm$^{-2}$ and
temperature $T=6.3^{+2.2}_{-1.5}$ keV, typical
for open clusters \citep[e.g.,][]{2019ApJ...871..116S}. 
$N_{\rm H}$ is about three times lower than the value obtained for \psr\
and consistent with the smaller distance to the cluster.
Further studies are needed to verify whether G18.45$-$0.42 is associated with Bica~3 or \psr.      

Star clusters can be sources of TeV emission \citep[e.g.][]{bednarek2007}.
However, currently only two open clusters are known to be associated with TeV 
sources\footnote{According to the catalog for Very High Energy Gamma-Ray Astronomy \citep{tevcat1}; \url{http://tevcat.uchicago.edu/}. The association
of HESS J1848$-$018 with the star-forming region W43 
is in question and it has been also considered 
as a PWN candidate \citep{acero2013}.}.
This makes the direct connection of HESS J1826$-$130 with Bica~3 unlikely. 
On the other hand, we can suggest that HESS J1826$-$130 is the TeV counterpart to
G18.45$-$0.42 basing on their positions and extents.


\section{Summary}
\label{sec:summary}

We have analysed the \xmm\ and \chan\ observations 
of the young $\gamma$-ray radio-quiet pulsar \psr\ and its PWN. 
The pulsar spectrum can be described by the PL model with a photon index $\Gamma\approx1$
and the PWN spectrum becomes softer with the distance from the pulsar.
We also estimated the distance to \psr\ to be $\approx$3.5 kpc, using the empirical relation 
between the distance and interstellar absorption by \citet{Marshall2006}.

\psr\ can be associated with the recently discovered SNR candidate G18.45$-$0.42.
This implies the pulsar transverse velocity of $\approx 530D_{3.5}$ km~s$^{-1}$, which is 
consistent with the pulsars velocity distribution, and the pressure driven snowplough 
phase of the remnant.

The Eel nebula appears to be a mixed-type morphology PWN
containing a torus, jets and a trail.
One of the jets is bent by the ram pressure, due to the pulsar \pmo\ vector 
not coinciding with the jet direction.
Such geometry explains the PWN morphology and supports the association
with G18.45$-$0.42.

The TeV source HESS J1826$-$130 overlaps with the \psr+PWN system
as well as with G18.45$-$0.42 and the open star cluster Bica~3.
Comparing the Eel X-ray and HESS J1826$-$130 $\gamma$-ray luminosities
with those of other X-$\gamma$-ray PWNe suggests
the TeV source is the relic PWN of \psr.
However, we cannot exclude the possibility that 
G18.45$-$0.42 can be the HESS J1826$-$130 counterpart
(or partially contribute to TeV emission).

Based on the spatial coincidence of G18.45$-$0.42 and Bica~3,
the latter can be the birthplace for the presumed SNR. 
The distance estimate to the open cluster is significantly lower than the distance to \psr\
which is confirmed by the X-ray spectral analysis. 
In this case G18.45$-$0.42 cannot be associated with the pulsar.

The pulsar \pmo\ measurement is necessary to solve the question 
about the Eel nebula's morphological type.
Together with confirmation of the G18.45$-$0.42 SNR nature
this can help to understand the relations between the pulsar, G18.45$-$0.42, HESS J1826$-$130 and Bica~3.

After this paper submission, the work by \citet{duvidovich2019} was published.
Authors used the same \xmm\ data to analyse the \psr\ and Eel emission.
They also found that the spectra of both objects can be described by power laws 
and the PWN spectrum softens with increasing distance from \psr.
They also argued that HESS J1826$-$130 is likely produced by Eel.
However, they did not use \chan\ data and performed spectral analysis
of only the brightest part of  Eel. There are no new distance 
constraint in their work and 
they did not  discuss the compact nebula morphology and
the connections between \psr, G18.45$-$0.42, HESS J1826$-$130 and Bica~3.
\section*{Acknowledgements}

We thank the anonymous referee for useful comments. 
The work of AVK and DAZ was supported by RF Presidential Programme MK$-$2566.2017.2. 
DAZ thanks Pirinem School of Theoretical Physics for hospitality.
The work of YAS was supported by the Fundamental Research Program of Presidium of the RAS P-12.
The scientific results reported in this article are based 
on data obtained from the \chan\ Data Archive and
observations obtained with \textit{XMM-Newton}, an ESA science mission 
with instruments and contributions directly funded by ESA Member States and the USA (NASA).
 



\bibliographystyle{mnras}

\bibliography{j1826} 

\bsp	
\label{lastpage}
\end{document}